\documentclass[12pt]{article}
\usepackage{euscript,amsmath,amssymb,latexsym}

\newcommand{\be}{\begin{equation}}
\newcommand{\ee}{\end{equation}}
\newcommand{\bea}{\begin{eqnarray}}
\newcommand{\eea}{\end{eqnarray}}

\begin{document}
\title{\bf
\vspace{1cm}Time Machine at the LHC}
\author{
I.Ya. Aref'eva and I.V. Volovich
 \\
{\it  Steklov Mathematical Institute}
\\ {\it Gubkin St.8, 119991 Moscow, Russia}\\
arefeva@mi.ras.ru;  volovich@mi.ras.ru}

\date {~}
\maketitle

\begin{abstract}
 Recently, black hole and brane production at
CERN's  Large Hadron Collider (LHC) has been widely discussed. We suggest that there is a
possibility to test causality at the LHC. We argue that if the scale of quantum gravity
is of the order of few TeVs, proton-proton collisions at the LHC could lead to the
formation of time machines (spacetime regions with closed timelike curves) which violate
causality. One model for the time machine is a traversable wormhole. We argue that the
traversable wormhole production cross section at the LHC is of the same order as the
cross section for the black hole production. Traversable wormholes assume violation of
the null energy condition (NEC) and an exotic matter similar to the dark energy is
required.
 Decay of the wormholes/time machines and signatures of time machine
events at the LHC are   discussed.
\end{abstract}
\maketitle

\newpage

$~~~~~~~~~~~~~~~~~~~~~~~~~~~~~~~~~~~~~~~~~~~~~~~~~~~~~~~$ {\it "It's against reason,"
said Filby.}

$~~~~~~~~~~~~~~~~~~~~~~~~~~~~~~~~~~~~~~~~~~~~~~~~~~~~~~~$ {\it "What reason?" said the
Time Traveller.}

$~~~~~~~~~~~~~~~~~~~~~~~~~~~~~~~~~~~~~~~~~~~~~~~~~~~~~~~~~~~~~~~~~$ {\it H.G. Wells, "The
Time Machine"}
\section{Introduction}

Causality is one of fundamental physical principles. We suggest in this note that there
is a possibility to test causality in experiments   at CERN's Large Hadron Collider
(LHC). This is related with a possibility of wormhole production in proton-proton
collisions at the LHC. The wormholes contain small spacetime regions with closed timelike
curves (CTC) which violate the standard causality condition.

A possibility of production in ultra-relativistic particle collisions of some objects
related with a non-trivial space-time structure is one of long-standing theoretical
questions. One of such particular objects is a black hole. Gravitational radiation in
collision of two classical ultrarelativistic black holes was considered by   D'Eath and
Payne \cite{DEa} and the mass of the assumed final black hole is estimated.

In general relativity there is Thorn's  hoop  conjecture which says that black holes form
when, and only when, a mass $M$ gets compacted into a region whose circumference  in
every direction is $\mathcal{C} < 4\pi GM$ \cite{Tho}. The area of the corresponding disk
is
\be
\label{k0} \pi r_0^2=4\pi G^2M^2\sim s/M_{\rm{Pl}}^4\,,
\ee
which gives a rough
estimate for the classical geometrical cross-section for black hole production. Here $G$
is the Newton constant, $M_{\rm{Pl}}=1/\sqrt{G}$ is the Planck mass and $s$ is the square
of the center of mass energy of colliding particles\,.

A conjecture that in string theory and in quantum gravity at energies much higher than
the Planck mass black hole
 production  emerges  has been made in \cite{ACV,tHo}.
 It has been proposed to use the Aichelburg-Sexl
shock wave metrics  to describe ultra-relativistic particles. Under collision of these
waves one can expect a production of black holes.

To speak on the production of black holes in quantum theory one should have a notion of a
quantum black hole as a state (pure or mixed) in some Hilbert space. We have to compute
the transition amplitude from a quantum state describing two particles to a quantum state
describing quantum black holes. A quantum gravity approach to this problem is discussed
in \cite{AVV}.  One considers the kernel of the
 transition amplitude
\be \label{k1} \langle h^{\prime\prime},\phi^{\prime\prime},\Sigma^{\prime\prime}
|h^{\prime},\phi^{\prime},\Sigma^{\prime}\rangle=\int \exp
\{\frac{i}{\hbar}S[g,\Phi]\}\mathcal{D}g\mathcal{D}\Phi \ee
 between
configurations of the three-metric $h_{ij}^{\prime}$ and fields $\phi^{\prime}$ on an
initial spacelike surface $\Sigma^{\prime}$ and a configuration $h_{ij}^{\prime\prime}$
and $\phi^{\prime\prime}$ on a final surface $\Sigma^{\prime\prime}$. In (\ref{k1}) the
integral is over all four-geometries $g_{\mu\nu}$, including summation over different
topologies, and field configurations $\Phi$, which match given values on the space-like
surfaces $\Sigma^{\prime}$ and $\Sigma^{\prime\prime}$, i.e. $\Phi
|_{\Sigma^{\prime}}=\phi^{\prime},~g |_{\Sigma^{\prime}}=h^{\prime}$ and $\Phi
|_{\Sigma^{\prime\prime}}=\phi^{\prime\prime},~g
|_{\Sigma^{\prime\prime}}=h^{\prime\prime}$.
 This formula assumes
the Wheeler-de Witt formalism \cite{Whe}, for a recent review see
\cite{GK}.

To get the transition amplitude between two particles
 and a black hole, or a wormhole one has to integrate the kernel (\ref{k1}) with
 the wave function $\Psi_{\Sigma^{\prime}} [h^{\prime},\phi^{\prime}]$ describing
 two particles and the wave function $\Psi_{\Sigma^{\prime\prime}}
 [h^{\prime\prime},\phi^{\prime\prime}]$  describing black
 hole or wormhole. An expression for the wave function of the ground
 state of a black hole is considered in \cite{BFZ}.

In the case of a semiclassical description of black holes production from particles
\cite{AVV} a leading contribution comes from $\Sigma^{\prime}$ being a partial Cauchy
surface with asymptotically simple past in a strongly asymptotically predictable
space-time and $\Sigma^{\prime\prime}$ being a partial Cauchy surface containing black
hole(s), i.e. $\Sigma^{\prime\prime}-J^-(\mathcal{T}^+)$ is non empty where
$J^-(\mathcal{T}^+)$ is the causal past of future null infinity, see \cite{HE}.

 A possible scenario for creation of black holes  by using classical solutions
 of the Einstein equations
 has been proposed in \cite{AVV}.
In this scenario it is supposed that ultra-relativistic particles
are represented by plane gravitational waves,  which interacting
collide and produce a black hole. A duality between plane
gravitational waves and black holes is used. Trans-Planckian
collisions in  standard quantum gravity have inaccessible energy
scale and cannot be realized in usual conditions. However if the
fundamental Planck scale of quantum gravity is of the order of few
TeVs \cite{AH} then one can argue that there is an exciting
possibility of production of black holes,
 branes, and Kaluza-Klein modes from the extra
 dimensions
in proton-proton collisions at CERN's Large Hadron Collider (LHC)
\cite{GT,DL,BF,Traped,BHP-review}. The cross section for creation of
a black hole or brane with radius $r_0$ was postulated to be
approximately equal to the geometrical cross section $\pi r_0^2$
\cite{BF} as in the hoop conjecture (\ref{k0}). The Schwarzschild
radius of a $4+n$ dimensional black hole of mass $M=\sqrt{s}$  is
approximately,
\be
\label{k3} r_0\sim
M_{4+n}^{-1}(s/M_{4+n}^2)^{\frac{1}{2(n+1)}}\,.
\ee Here $M_{4+n}$
is the $4+n$ dimensional Planck mass and the 4 dimensional Planck
mass is given by
\be
\label{k4} M_{\rm{Pl}}^2\sim
V_nM_{4+n}^{2+n},
\ee
where $V_n$ is the volume of the extra
dimensions.

 This process can be achieved by scattering of two partons with the
center of mass energy $\sqrt{s}$ larger than $M$ and impact
parameter smaller than $r_0$. For a discussion of different
viewpoints see \cite{Vol,Ran}.

D'Eath and Payne \cite{DEa} have studied the problem of classical
collision with zero impact parameter and shown that a closed trapped
surface forms. This analysis was extended to a nonzero impact
parameter by Eardley and Giddings \cite{Traped}. The Aichelburg-Sexl
solution has the form \be\label{AS} ds^2=-dudv+dx^{i2}+\varphi
(x^i)\delta (u)du^2, \ee where $\varphi$ depends only on the
transverse coordinates $x^i$. A marginally trapped surface is
constructed in the union of two incoming null hypersurfaces by
solving a constraint problem for the Dirichlet Green's function.

In this note we consider a possibility of production of time machines at the LHC. In
general relativity a timelike curve in space-time represents a possible path of an object
or an observer. Normally such a curve will run from past to future, but in some
space-times the curves can intersect themselves, giving a closed timelike curve (CTC)
which is interpreted as a time machine. It suggests the possibility of time-travel with
its well known paradoxes.

There are many solutions of the Einstein equations  with CTCs. A list of such solutions
includes G${\rm\ddot{o}}$del's solution \cite{God}, van Stockum and Tipler cylinders
\cite{Tip}, Kerr and Kerr-Newman solutions \cite{HE}, Gott's time machine \cite{Got},
Wheeler wormholes (space-time foam) \cite{Whe},  Morris-Thorne traversable wormholes
\cite{MT}, and Ori's dust asymptotically-flat space-time \cite{Ori}, see \cite{Vis1,LLO}
for a review. Chronology protection in AdS/CFT is considered in \cite{Her}.
G${\rm\ddot{o}}$del universes also appear
 in string theory and they are T-dual
 to $pp$-waves
 \cite{GH}.
Euclidean wormholes
 are discussed in \cite{Arkani-wh,Sundrum}. Higher dimensional
 wormholes are considered in \cite{DD,AP}.

A wormhole forms a handle-like geometry, whose two mouths join different regions of
spacetime. If the wormhole is traversed from mouth to mouth, it acts as a time machine
allowing one to travel into the past or into the future.

Violation of normal chronology is so objectionable an occurrence
that any such solution could be rejected as unphysical. However, the
Einstein equations are local equations and therefore one has to
impose additional principles to preserve chronology. There are long
debate concerning such principles
\cite{Tip,Vis1,DJH,Haw,Vis3,Wal,Kay}. In particular, in \cite{DJH}
it was shown that acausal CTC in Gott`s universe cannot be realized
by physical, timelike, sources.

An attempt to save causality and exclude CTCs from general
relativity is Hawking's "chronology protection conjecture" which
asserts that the law of physics do not allow the appearance of CTC
\cite{Haw}. However, there are not enough convincing arguments for
this conjecture. Indeed, it was suggested that divergences in the
energy-momentum tensor occur when one has closed causal curves.
These divergences may create spacetime singularities which prevent
one from traveling through to the region of closed timelike curves.
However, it might be that quantum gravitational effects may smear
out the divergences. Moreover, if one believes that there exists a
full theory of quantum gravity, then chronology protection should be
settled by using this theory \cite{Vis3}.

Whether the chronology protection conjecture can be derived from the known physics laws
or it is an independent postulate  is  still an open question. In this note we suggest to
test it in experiments at the LHC.

Note that  the CTC problem probably is related with the
irreversibility problem well known in statistical physics. For a
discussion of the black hole information paradox see \cite{NV} where
it is explained  that the black hole information paradox is a particular
case of the irreversibility problem which is not solved not only for
black {\it hole} but even for the usual black {\it body}.

\section{Traversable Wormholes and NEC}

The four-dimensional spacetime metric representing a spherically
symmetric and static wormhole is given by \cite{MT,Vis1}
\be
\label{t1} ds^2=- e^{2\Phi
(r)}dt^2+\frac{dr^2}{1-f(r)/r}+r^2(d\theta^2+\sin^2\theta
d\phi^2)\,.
\ee
Here $\Phi (r)$ is designated the redshift function
and $f(r)$ is denominated the shape function. The radial coordinate
$r$ varies from $r =r_0$ corresponding to the wormhole throat,
$f(r_0)=r_0$, to  some  $R$. The redshift function supposed to be
finite, i.e. the event horizon is absent for $r_o< r< R$ and the
shape function should satisfy the following inequality $f'r-f<0$.
For asymptotically flat wormholes $R=\infty$. Wormholes with a
cosmological constant are considered in \cite{LLO}.

As it is well known  traversable wormholes exist only for  NEC
violating stress energy tensors \cite{MT}. According to the NEC
\cite{HE} the stress energy tensor $T_{\mu\nu}$ has to satisfy the
requirement $T_{\mu\nu}k^\mu k^\nu \geq 0$, where $k^\mu $ is a null
vector, $k^\nu k_\nu=0$. Using the Einstein field equations,
$G_{\mu\nu}=M_{\rm{Pl}}^{-1} T_{\mu\nu}$, one obtains \cite{MT} the
following expression for the sum of the energy density $\rho(r)$ and
the radial pressure $p_r(r)$
$$\rho(r)+p_r(r)=\frac{1}{M_{\rm{Pl}}}
\left(\frac{f'r-f}{r^3}+2\left(1-\frac{f}{r}\right)
\frac{\Phi'}{r}\right)\,.$$ We see that the embedding condition
together with the requirement of finiteness of  the redshift
function  lead  to the NEC violation on the wormhole throat.

Several scenario of the NEC violating have been   considered in recent years. Generally
speaking  the NEC violating means unstability. But this is true only under special
assumptions. There are examples of stable effective theories where the NEC is violated
\cite{DGNR}. In these particular cases the Lorentz invariance is broken and superluminal
modes are present. Typical features of NEC violating effective theories is  a presence of
higher derivative terms and also superluminal modes.  Gravitational Lorentz violation and
superluminality take place also for wormhole solutions in Euclidean AdS gravity
\cite{Sundrum}. Note that traversable wormholes may be also supported  by the dark energy
(see for example \cite{Are,AV} and refs therein) with the equation of state parameter
$w<-1$ \cite{Sushkov}.

In the brane world scenario, where the Universe is considered as a 3-brane embedded in a
D-dimensional bulk, the four-dimensional Einstein field equations contain the effective
four-dimensional stress energy tensor,
 \be
 G_{\mu\nu}= M_{\rm{Pl}}^{-1}T^{\rm{\rm{eff}}}_{\mu\nu}.
 \ee
$T^{\rm{eff}}_{\mu\nu}$ is a sum of the  stress energy tensor of a matter confined
 on the brane, $T_{\mu\nu}$
 and correction terms that arise from a projection of the  D-dimensional
 Einstein equation
 to the four-dimensional space-time.
 It is possible that $ T^{\rm{eff}}_{\mu\nu}$ supported the four-dimensional
whormhole solution
 violates the NEC meanwhile $T_{\mu\nu}$ does not violate the NEC.

 In the  simplest brane world scenario where the  Universe is considered
as a 3-brane embedded in a five-dimensional bulk   these correction
terms can be written explicitly \cite{Maeda,Aliev},
\bea
T^{\rm{eff}}_{\mu\nu}&=&T_{\mu\nu}+\frac{6}{M_4\lambda }
\Pi _{\mu\nu}-E _{\mu\nu},\\
\Pi _{\mu\nu}&=&\frac{1}{12}T\,T_{\mu\nu}- \frac{1}{4}T_{\mu\alpha}\,T^\alpha_{\nu}+
\frac{1}{8}g_{\mu\nu}[T_{\alpha\beta}\,T^{\alpha \beta}-
\frac{1}{3}T^2],\\
E _{\mu\nu}&=&^{(5)}C_{\mu \alpha \nu \beta}n^\alpha n^\beta, \eea where $\,^{(5)}C_{\mu
\alpha \nu \beta}$ is the five-dimensional Weyl tensor, $\alpha,\beta=0,1,2,3,4$ and
$n^\alpha$ is the unit normal to the brane. These formulas give the following relation
between $\rho^{\rm{eff}}+p_r^{\rm{eff}}$ and $\rho+p_r$ \be
\rho^{\rm{eff}}+p_r^{\rm{eff}}=\rho+p_r-\frac{1}{8\pi}(\epsilon+\sigma_r)+
\frac{1}{\lambda}\rho(\rho+p_r)\,. \ee Here $\rho(r)$ and $p_r(r)$ are the energy density
and the radial pressure of the matter confined on the brane, $\epsilon$ and $\sigma_r$
are diagonal components of the projected Weyl tensor $diag[\epsilon(r),\sigma_r(r),$
$\sigma_t(r),\sigma_t(r)]$. Now to have a wormhole one has to provide the condition \be
8\pi(\rho+p_r)(1+\frac{\rho}{T})<\epsilon+\sigma_r\,. \ee As comparing with
four-dimensional wormholes we see a softening of the energy condition. This relaxed
condition appears due to corrections from the Weyl tensor in the bulk (compare with the
NEC violation from the string field non-local action \cite{Are,AV}). For some particular
examples it is possible to show that the four-dimensional effective stress energy tensor
violates the NEC meanwhile the total five-dimensional stress energy tensor does respect
the NEC \cite{AP}. We do not present here higher-dimensional solutions corresponding to
wormholes on 3-brane. This is a subject of recent studies, see ref.\cite{DD,AP}.
It would be interesting to find the wormhole solutions in the context of intersecting D5-branes related with the
Standard model
\cite{CIM}. For a
general class of solutions one expects the following dependence of the radius of the
throat or mouth $r_0$ from the mass \be
r_0=\gamma_{\rm{wh}}(D)\frac{1}{M_D}\left(\frac{M_{\rm{wh}}} {M_D}\right)^{\alpha}\,. \ee
This formula is similar to the formula for the  Schwarzschild radius for the
D-dimensional Schwarzschild solution: \be
ds^2=-(1-(\frac{r_s}{r})^{D-3})dt^2+(1-(\frac{r_s}{r})^{D-3})^{-1}
dr^2+r^2d\Omega^2_{D-2}\,, \ee where the Schwarzschild radius $r_s$ is related to the
mass of the black hole  by the relation \be r_s=\gamma_{\,_{\rm{bh}}}(D)\frac{1}{M_D}
\left(\frac{M_{\,_{\rm{bh}}}}{M_D}\right)^{\alpha_{\,_{\rm{bh}}}},\,\,\,\,\,
\alpha_{\,_{\rm{bh}}}= \frac1{D-3}\,, \ee where $\gamma_{\,_{\rm{bh}}}(D)=1/\sqrt{\pi}
(\frac{8\Gamma(D-1/2)}{D-2})^{1/(D-3)}$. Let us note that for solution (\ref{t1}) the
radius of the throat $r_0$ is larger than the Schwarzschild  radius: $r_0>r_s$\,.

\section{Wormhole Production at Accelerators} To compute the wormhole production cross
section we can follow the approach for computation of the black hole production cross
section \cite{GT,DL,BF,Traped,BHP-review}. The wormhole cross section is found from the partonic
cross section for partons $i$ and $j$ to form a wormhole: \be \label{w1} \sigma_{pp\to
\rm{wh}}(s)\sim \sum_{ij}\int_{\tau_m}^1d\tau
\int_{\tau}^1\frac{dx}{x}f_i(x)f_j(\tau/x)\sigma_{ij\to \rm{wh}} (\tau s)\,. \ee Here
$\sqrt {s}$ is the  center of mass energy, $x$ and $\tau/x$ are the parton momentum
fractions, and $f_i$ are the parton distribution functions. The parameter
$\tau_m=M^2_{min}/s$ where $M_{min}$ corresponds to the minimum mass for a valid wormhole
description. $f_i$ are the Parton Distribution Functions (PDFs), (we suppress here
transfer momenta). This formula is the same as the formula for black hole production, the
difference being only in numerical factors.

The geometrical cross section of the wormhole production is \be\label{2} \sigma_{ij\to
\rm{wh}} (s)=\pi F(\sqrt{s}/M_D) r_0^2 (\sqrt{s},M_D)\,. \ee
 The form factor $F(\sqrt{s}/M_D)$ incorporates  the theoretical
uncertainties in description of the process, such as the amount of the initial center
mass energy that goes into the wormhole, the distribution of wormhole masses as function
of energy. These  corrections are similar to corrections in the formula for black hole
production.

\section{Conclusions}

Causality is the fundamental physical principle. In quantum
field theory causality
 and the spacetime picture of the high energy
scattering  were considered in \cite{BS,GIP,BVT}. If there are spacetime
regions with CTC (time machines) then causality is violated.

In this note we suggested to test causality by using experiments at
the LHC. We argued that if one can trust the classical geometrical
estimate of the cross-section for the black hole production, if
there exists an exotic matter similar to the dark energy, and if the
scale of quantum gravity is of the order of few TeVs then one can
expect the production of time machines/wormholes in the
proton-proton collisions at the LHC of the same order as the cross
section for the black hole production. This would leads to violation
of the standard causality condition. Further studies of the
experimental signatures of the wormhole production
 are required
since there are transitions between black holes and wormholes \cite{Hay}.

It would be interesting to explore in some details the formula
(\ref{k1}) for the transition amplitude between colliding quantum
particles and black holes/wormholes which should be integrated with
the wave function of the wormholes.

\section*{Acknowledgements}

We are grateful to D. Astefanesei, V. Cardoso, S. Deser, B.
Dragovich, V.P. Frolov, R. Jackiw, G. 't Hooft, A. Mironov  for
critical comments, useful discussions and correspondence. The work
of I.A. and I.V. is supported in part by INTAS grant 03-51-6346.
I.A. is also partially supported by RFBR grant 05-01-00758  and
Russian President's grant NSh-2052.2003.1 and I.V. is also partially
supported  by RFBR grant 05-01-00884  and Russian President's grant
NSh-1542.2003.1.

\section*{Note Added}

After submission to the arXiv of the first version of this paper,
there appeared a paper\cite{MMT} discussing possible observable
traces of mini-time-machines. It seems the important question on
possible experimental signatures of time machines deserves further
explorations.

 \end{document}